\documentclass[pra,aps,twocolumn,superscriptaddress,showpacs,showkeys]{revtex4-1}
\usepackage{amsmath,amsthm,amssymb,graphicx,subfigure,xcolor}
\usepackage{graphics,float}
\usepackage{color}
\usepackage[unicode=true]{hyperref}
\hypersetup{
     colorlinks=true,       		
     linkcolor=blue,          	
     citecolor=red,            
     urlcolor=magenta,           	
 }

\begin{document}

\title{Chained Einstein-Podolsky-Rosen steering inequalities with improved visibility}

\author{Hui-Xian Meng}
\affiliation{Theoretical Physics Division, Chern Institute of Mathematics, Nankai University, Tianjin 300071, People's Republic of China}

\author{Jie Zhou}
\affiliation{Theoretical Physics Division, Chern Institute of Mathematics, Nankai University, Tianjin 300071, People's Republic of China}

\author{Changliang Ren}
\affiliation{Center for Nanofabrication and System Integration, Chongqing Institute of Green and Intelligent Technology, Chinese Academy of Sciences, People's Republic of China}
\affiliation{Chongqing Key Laboratory of Multi-Scale Manufacturing Technology, Chongqing Institute of Green and Intelligent Technology, Chinese Academy of Sciences, People's Republic of China}

\author{Hong-Yi Su}
\email{hysu@mail.nankai.edu.cn}
\affiliation{Theoretical Physics Division, Chern Institute of Mathematics, Nankai University, Tianjin 300071, People's Republic of China}

\author{Jing-Ling Chen}
\email{chenjl@nankai.edu.cn}
\affiliation{Theoretical Physics Division, Chern Institute of Mathematics, Nankai University,
 Tianjin 300071, People's Republic of China}
\affiliation{Centre for Quantum Technologies, National University of Singapore,
 3 Science Drive 2, Singapore 117543}

\date{\today}
\begin{abstract}
It is known that the linear $n$-setting steering inequalities introduced in Ref. [Nature
Phys.  {\bf{6}}, 845 (2010)] are very efficient inequalities in detecting steerability of the Werner states by using optimal measurement axes. Here, we construct \emph{chained} steering inequalities that have improved visibility for the Werner states under a finite number of settings. Specifically, the threshold values of quantum violation of our inequalities for the $n=4,6,10$ settings are lower than those of the linear steering inequalities. Furthermore, for almost all \emph{generalized} Werner states, the chained steering inequalities always have improved visibility in comparison with the linear steering inequalities.
\end{abstract}

\pacs{03.65.Ud, 03.67.Mn, 42.50.Xa}
\keywords{Linear n-setting steering inequalities; Chained Einstein-Podolsky-Rosen steering inequalities; Threshold values of quantum violation}
\maketitle

\section{Introduction}

Entanglement is a property of distributed quantum
systems that has no classical counterpart and
challenges our everyday-life intuition about the physical
world~\cite{Horodecki}. It is  the key element in many quantum
information processing tasks~\cite{Nielsen}.
The origin of steering, a  notion closely related with entanglement,
 can
historically be traced back to Schr$\rm{\ddot{o}}$dinger's reply~\cite{Schrodinger} to the well-known Einstein-Podolsky-Rosen (EPR) argument~\cite{epr35}.
However, steering lacked operational meanings, until  the year 2007 when it was given
 a rigorous definition through the quantum information task~\cite{Wiseman}. Then it  becomes clearer
that the EPR paradox concerns more precisely the existence of local hidden state (LHS) models, rather than that
of local hidden variable (LHV) models~\cite{Bell,epr35,gisin91,wernerJPA}. After 2007, EPR steering has gained a very rapid development in theory \cite{Chen,Skrzypczyk,Bowles,chenSR,chenRC,Cavalcanti} and experiment \cite{wisemanNature,Sun}.
Different from entanglement and Bell nonlocality~\cite{Brunner}, steering possesses a curious feature of ``one-way quantumness''~\cite{Bowles}, which is
shown useful in one-way quantum information tasks like, e.g., one-way quantum cryptography~\cite{one-way-steer}.

Intriguingly, the authors of~\cite{Wiseman}  proved that steerable states are a strict subset
of the entangled states, and a strict superset of the states that can exhibit Bell nonlocality.
In analogy to the violation of Bell's inequality which implies Bell nonlocality, any violation of a steering inequality immediately implies EPR steering~\cite{chenRC,wisemanNature}.
It is in principle easier to experimentally observe the violation than Bell inequalities,
because  one has no concerns about closing the notorious locality loophole as in a Bell test~\cite{Brunner}.

Among many others, the $n$-setting linear steering inequalities in~\cite{wisemanNature} have been experimentally verified by the violation with the Werner states in their optimal measurement axes. In this paper, comparing with those in~\cite{wisemanNature}, we focus on the generalized Werner states and address the question of whether there exist steering inequalities that have improved visibility so that steering is more robust against white noise and thus more experiment-friendly to test in practice. Below, in the Alice-to-Bob steering scenario, we shall construct chained steering inequalities out of the chained Bell inequalities, and demonstrate that with Bob's measurement directions properly adjusted, one can obtain \emph{lower} threshold values above which these inequalities can be violated.


The paper is structured as follows. In Sec.~\ref{1}, we shall briefly review two family of $n$-setting steering inequalities: the linear~\cite{wisemanNature} and the chained~\cite{Meng}, along with threshold values of their violation. In Sec.~\ref{2}, we will present new measure directions, obtain new chained steering inequalities, and then compute the threshold values for the obtained chained inequalities, which are shown to be lower than those of the linear inequalities. In Sec.~\ref{BI}, focus on the generalized Werner states, we   research the region of their parameters that can have the inequalities violated, and  find that for almost all generalized Werner states the new chained steering inequalities always have improved visibility in comparison with the linear steering inequalities. We discuss the results in Sec.~\ref{conclusion}.

\section{Two known families of steering inequalities}\label{1}

In this section, we first review two families of steering inequalities, and then give  the threshold values of their violation.

\subsection{The $n$-setting linear steering inequalities}

For qubit systems, we can take Bob's $i$-th measurement as the Pauli observable $\hat{\sigma}_i^B$ along some axis $b[i]$. By denoting Alice's declared result(we make no assumption that it is derived from a quantum measurement) as random variables
$A_i\in\{1,-1\}$ for all $i$, the $n$-setting linear steering inequality is of the form
\begin{align}\label{eq18}
I_n\equiv\frac{1}{n}\sum_{i=1}^n\langle A_i\hat{\sigma}_i^B\rangle\overset{\rm LHS}{\leq} C_n.
\end{align}
The LHS bound $C_n$ is the maximum value that $I_n$ can reach if Bob has a pre-existing state known to Alice, rather than half of an entangled pair shared with Alice.
Clearly, the quantum bound of $I_n$ is just $1$.
In general, for a given set of Bob's directions
\begin{equation}\label{eq3}b[i]=\left(\sin\theta_{b_i}\cos\phi_{b_i}, \sin\theta_{b_i}\sin\phi_{b_i},
 \cos\theta_{b_i}\right),\end{equation}
one has then
 \begin{eqnarray*}
\hat{\sigma}_i^B=\sin\theta_{b_i}\cos\phi_{b_i}\sigma_x+\sin\theta_{b_i}\sin\phi_{b_i}\sigma_y+
\cos\theta_{b_i}\sigma_z,
\end{eqnarray*}
where $\sigma_x,\sigma_y,\sigma_z$ are the Pauli matrices.
If the directions are chosen as along the vertices of Platonic solids, as in Figure 2 of Ref. \cite{wisemanNature}, the  analytical expressions for the bounds read
\begin{equation}\label{eq19}
\begin{aligned}
&C_2=\frac{1}{\sqrt{2}}\approx 0.707107,\\
&C_3=C_4=\frac{1}{\sqrt{3}}\approx 0.57735,\\
&C_6=1-\frac{5}{3\sqrt{10+2\sqrt{5}}}\sqrt{4-\sec^2{\frac{3\pi}{10}}}\\
&~~~=\frac{1+\sqrt{5}}{6}\approx 0.539345,\\
&C_{10}=1-\frac{1}{10}\left(1+\frac{\tan{\frac{2\pi}{5}}}{\sin{\frac{\pi}{5}}}\right)\sqrt{\frac{24}{3+\sqrt{5}}-4}\\
&~~~=\frac{3+\sqrt{5}}{10}\approx 0.523607.
\end{aligned}\end{equation}


\subsection{The $n$-setting chained steering inequalities}\label{BI}

The chained steering inequality for a given $n$
is of the form~\cite{Meng}
\begin{equation}\label{eq1}
\mathcal{I}_n\overset{{\rm LHS}}{\leq}\mathcal{C}_n,
\end{equation}
where
\begin{equation}\label{eq2}
\begin{aligned}
&\mathcal{I}_n=  \langle A_1\hat{\sigma}_1^B \rangle+  \langle A_2\hat{\sigma}_1^B\rangle+ \langle A_2 \hat{\sigma}_2^B \rangle+\\
&~~~~~~~~~~~  \langle A_3\hat{\sigma}_2^B \rangle+  \langle A_3\hat{\sigma}_3^B \rangle+  \langle A_4\hat{\sigma}_3^B \rangle+\cdots+\\
&~~~~~~~~~~~  \langle A_{n}\hat{\sigma}_n^B\rangle-  \langle A_1\hat{\sigma}_n^B\rangle,
\end{aligned}
\end{equation}
and
 $\mathcal{C}_n$
is the bound for LHS models.
For each $n$, if Bob chooses his measurement directions as
\begin{equation}\label{eq5}
b[i]=\left(\sin{\frac{(2i-1)\pi}{2 n}}, 0, \cos{\frac{(2i-1)\pi}{2n}}\right),
\end{equation}
with $i=1,2,\cdots,n$,
then the LHS bound equals~\cite{Meng}
\begin{equation}\label{eq6}
  \mathcal{C}_n=2\cot{\frac{\pi}{2n}},
\end{equation}
which is attained at $A_1=A_2=\cdots=A_n=1.$

 \subsection{Threshold value of quantum violation}\label{BI}

 For a given set of Alice's measurement directions
\begin{equation}\label{eq8}a[i]=\left(\sin\theta_{a_i}\cos\phi_{a_i}, \sin\theta_{a_i}\sin\phi_{a_i},
 \cos\theta_{a_i}\right),\end{equation}
the observables that Alice uses read
 \begin{eqnarray*}
A_i=\sin\theta_{a_i}\cos\phi_{a_i}\sigma_x+\sin\theta_{a_i}\sin\phi_{a_i}\sigma_y+
\cos\theta_{a_i}\sigma_z.
\end{eqnarray*}
The quantum violation, $\mathcal{Q}_n$, can then be obtained by numerating all possible $a[i]$.
%
%

Let us consider the Werner state
\begin{equation}\label{eq9}
\rho=V|\psi\rangle\langle\psi|+(1-V)\frac{I_4}{4},
\end{equation}
where $|\psi\rangle$ denotes the maximally entangled state
\begin{equation}\label{eq10}|\psi\rangle=\frac{1}{\sqrt{2}}(|0\rangle \otimes |1\rangle -|1\rangle \otimes |0\rangle),
\end{equation}
$|0\rangle$ and $|1\rangle$ are eigenstates of $\sigma_z$ with eigenvalues 1 and $-1$, respectively,
 and $I_4$ is the identity
matrix on  $\mathbb{C}^2\otimes\mathbb{C}^2$. Let us define the visibility as
\begin{equation}\label{eq12}
\mathcal{V}_n={\mathcal{C}_n}/{\mathcal{Q}_n}.
\end{equation}
For each $n$-measurement setting, the quantity $\mathcal{V}_n$ describes
the {\it{threshold value}}, above which the state (\ref{eq9}) cannot be described by LHS models.

\emph{Remark 1.---}The visibility for inequality (\ref{eq18}), defined similarly as
 \begin{equation}
   {V}_n={{C}_n}/{{Q}_n}={{C}_n}/{1}={C}_n,
 \end{equation}
 equals $V$ for all $n$. Namely, inequality (\ref{eq18}) can only detect by its violation the steerability of Werner states with
$V>C_n$.

\emph{Remark 2.---}For Bob's directions (\ref{eq5}), the maximum quantum violation is computed as \cite{Meng}
\begin{equation}\label{eq13}
\mathcal{Q}_n=2n\cos{\frac{\pi}{2n}},
\end{equation}
and so from (\ref{eq6}) and (\ref{eq12}) the visibility for (\ref{eq1}) equals
\begin{align}\label{eq14}
\mathcal{V}_n=\frac{1}{n\sin{\frac{\pi}{2n}}}.
\end{align}

The directions (\ref{eq5}) are not optimal for inequality (\ref{eq1}), as there exist other directions to obtain a lower visibility. In the following section, we will present new measurement directions, obtain new chained steering inequality. Moreover, we compute  threshold values for these new chained steering equalities and compare them with ones for the $n$-setting linear steering inequalities and (\ref{eq14}).

\section{Chained Steering Inequalities with improved visibility}\label{2}

\subsection{A simple example: $n=4$}

When Bob chooses
\begin{equation}
  \begin{split}
    &\theta_{b_1}=\frac{5\pi}{4},~\phi_{b_1}=\frac{\pi}{3},~\theta_{b_2}=0,~\phi_{b_2}=\frac{11\pi}{6},\\
    &\theta_{b_3}=\frac{3\pi}{5},~\phi_{b_3}=\frac{3\pi}{4},~\theta_{b_4}=\frac{3\pi}{2},~\phi_{b_4}=\frac{\pi}{2},
  \end{split}
\end{equation}
i.e.,
\begin{equation}
  \begin{split}
    &b[1]=\left(-\frac{\sqrt{2}}{4}, -\frac{\sqrt{6}}{4}, -\frac{\sqrt{2}}{2}\right),\\
    &b[2]=\biggr(0, 0, 1\biggr),\\
   &b[3]=\left(-\frac{\sqrt{5+\sqrt{5}}}{4}, \frac{\sqrt{5+\sqrt{5}}}{4},\frac{1-\sqrt{5}}{4}\right),\\
   &b[4]=\biggr(0, -1, 0\biggr),
  \end{split}
\end{equation}
one can have
\begin{equation}
  \begin{split}
    &\mathcal{Q}_4\approx3.630746,~~ \mathcal{C}_{4}\approx2.055877,
    ~~\mathcal{V}_4\approx0.566241.
  \end{split}
\end{equation}
Here, the visibility is lower than that of the linear steering inequality, for which ${V}_4=C_4\approx0.57735$. This example implies that there exists new chained steering  inequality such that  its visibility  lower than that of (\ref{eq18}). We shall  find  new chained steering inequalities  with lower threshold values numerically and list the results below.

\subsection{The lowest threshold values: $n=2,3,4,6,10$}

We now list Bob's measurement directions satisfying that for these directions the new chained steering inequalities  have the lowest threshold values compared with the $n$-setting linear steering inequality and the chained steering inequalities in~\cite{Meng}.

\begin{itemize}
  \item $n=2$: For arbitrary directions that Bob chooses, we always have
  \begin{equation}
    \mathcal{V}_2=\frac{1}{\sqrt{2}}\approx0.707107.
  \end{equation}

\item $n=3$: If Bob chooses the directions (\ref{eq5}), then
\begin{equation}
  \mathcal{Q}_3= 3\sqrt{3},~~~\mathcal{C}_3=2\sqrt{3},
\end{equation}
and so
\begin{equation}
\mathcal{V}_3=\frac{2}{3}\approx0.666667.
\end{equation}
In fact, this is the minimal threshold value, which can be confirmed by running all possible directions numerically.
What is worth noting is that to get this value Bob's directions
are not unique. For instance, if
Bob chooses instead the following
\begin{equation}\label{eq2114}
\begin{split}
b[1]=(0,0,1),\\
b[2]=(1,0,0),\\
b[3]=(0,1,0),
\end{split}
\end{equation}
i.e.,
\begin{equation}
  \begin{split}
    &\theta_1=0,~~\phi_1=0,~~\theta_2=\frac{\pi}{2},\\
    &\phi_2=0,~~\theta_3=\frac{\pi}{2},~~\phi_3=\frac{\pi}{2},
  \end{split}
\end{equation}
which is the same to what Bob chooses in the Figure 2 of Ref. \cite{wisemanNature},
then, again,
\begin{equation}\label{eq2117}
\begin{aligned}
&\mathcal{Q}_3= 3\sqrt{2},~~\mathcal{C}_3=2\sqrt{2},~~\mathcal{V}_3=\frac{2}{3}.\end{aligned}\end{equation}

\item $n=4$: Bob chooses
\begin{equation}\label{eq15}
\begin{split}
b[1]&=(0.387712, 0.325511, 0.862393),\\
b[2]&=(0.662244, 0.014395, -0.74915),\\
b[3]&=(-0.393555, -0.764746, 0.510174),\\
b[4]&=(0.256293, 0.92029, 0.295602),
\end{split}
\end{equation}
such that
\begin{equation}
  \begin{split}
    &\mathcal{Q}_4\approx 3.605552,~~\mathcal{C}_4=2,~~\mathcal{V}_4\approx0.5547.
  \end{split}
\end{equation}

\item $n=6$: Bob chooses
\begin{equation}\label{eq17}
\begin{split}
b[1]&=(-0.40343, -0.594926, -0.695203),\\
b[2]&=(0.786331, -0.613429, 0.073405),\\
b[3]&=(0.39261, 0.6444655, 0.656141),\\
b[4]&=(-0.445744, -0.539581, 0.714258),\\
b[5]&=(0.795316, -0.595208, 0.114891),\\
b[6]&=(0.435205, 0.494333, -0.752484),
\end{split}
\end{equation}
such that
\begin{equation}
  \begin{split}
    &\mathcal{Q}_6\approx 8.387765,~~~\mathcal{C}_6\approx  4.426295,\\
    &~~\mathcal{V}_6\approx0.527709.
  \end{split}
\end{equation}

\item $n=10$: Bob chooses
\begin{equation}\label{eq18-0}
\begin{split}
b[1]&=(-0.236305, -0.475636, 0.847308),\\
b[2]&=(-0.858832, 0.00901434, 0.512178),\\
b[3]&=(0.320136, 0.687104, 0.652228),\\
b[4]&=(-0.137947, 0.990267, -0.0185065),\\
b[5]&=(0.774103, 0.422827, -0.47115),\\
b[6]&=(0.637737, 0.18577, 0.747517),\\
b[7]&=(0.0326946, 0.0548708, 0.997958),\\
b[8]&=(0.829632, -0.326344, -0.453001),\\
b[9]&=(0.723801, -0.672939, 0.152531),\\
b[10]&=(-0.296751, -0.478204, 0.826595),
\end{split}
\end{equation}
such that
\begin{equation}
  \begin{split}
    &\mathcal{Q}_{10}\approx 14.702807,~~~\mathcal{C}_{10}\approx 7.615109,\\
     &~~\mathcal{V}_{10}\approx0.517936.
  \end{split}
\end{equation}

\end{itemize}

%

A comparison between the results on the linear inequalities (\ref{eq18}), on the chained steering inequalities in~\cite{Meng}, and the new chained steering inequalities  is listed in  Table~\ref{table2}.
\begin{table}[H]
\caption{\label{table2} Comparison of the visibility of inequality (\ref{eq18}), inequalities in~\cite{Meng}, and the new chained inequalities.}
\begin{ruledtabular}
\begin{tabular}{cccccc}
$V_n,~\mathcal{V}_{n}$ & $n=2$ & $n=3$ & $n=4$ & $n=6$ & $n=10$\\
  \hline
  Inequality (\ref{eq18}) & $\frac{1}{\sqrt{2}}$ & $\frac{1}{\sqrt{3}}$ & $\frac{1}{\sqrt{3}}$ & $\frac{1+\sqrt{5}}{6}$ & $\frac{3+\sqrt{5}}{10}$\\
Inequality in~\cite{Meng} & $\frac{1}{\sqrt{2}}$ & $\frac{2}{3}$ & $\frac{1}{4\sin{\frac{\pi}{8}}}$ & $\frac{1}{6\sin{\frac{\pi}{12}}}$ & $\frac{1}{10\sin{\frac{\pi}{20}}}$\\
 The new inequality &  $\frac{1}{\sqrt{2}}$ & $\frac{2}{3}$ & $0.5547$ & $0.527709$ & $0.517936$\\
\end{tabular}
\end{ruledtabular}
\end{table}

\begin{figure}[H]
  \centering{\includegraphics[width=75mm]{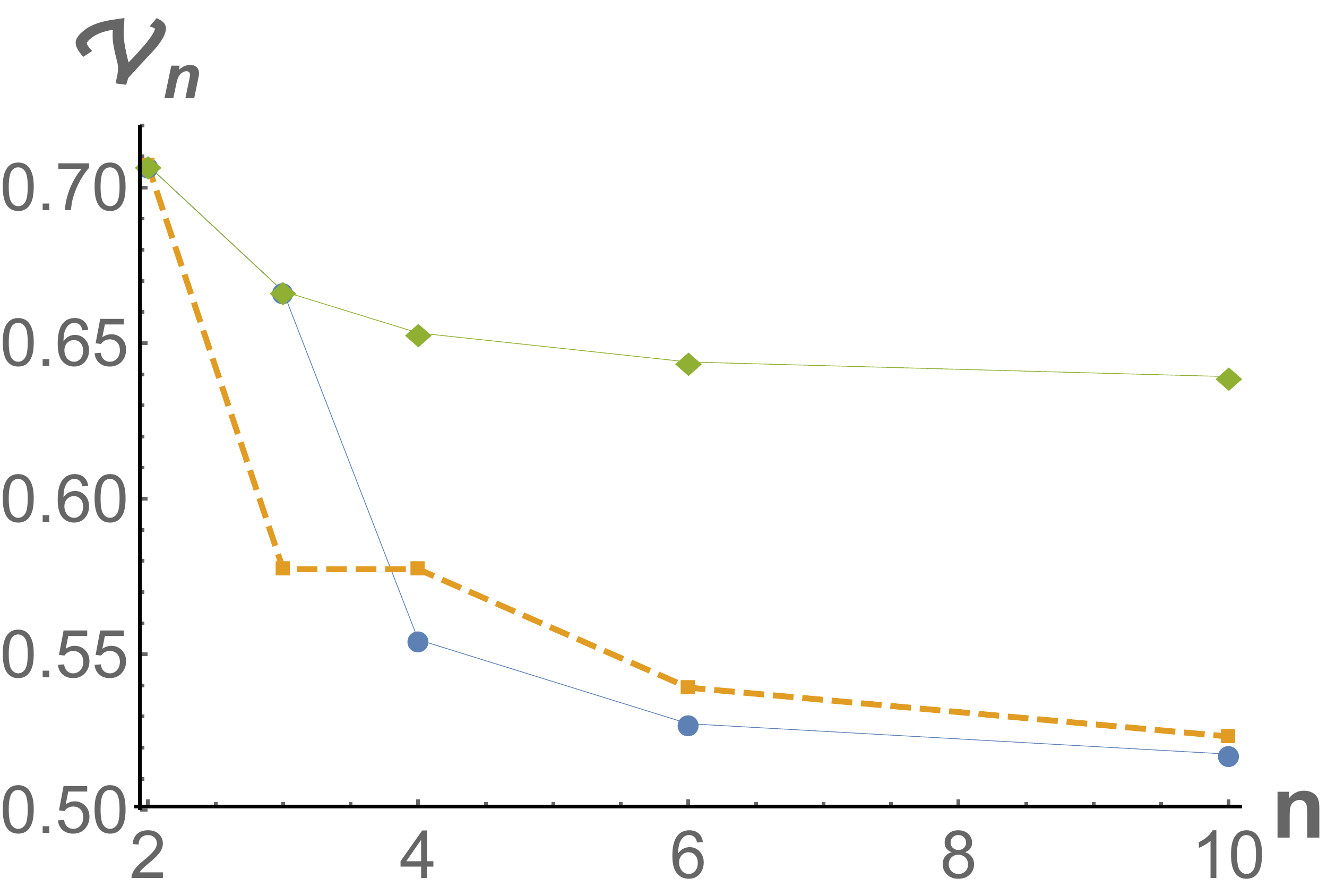}}
  \caption{\label{fig1}The relationship between the minimal $\mathcal{V}_{n}$ and $n$ for the inequality (\ref{eq18}), inequalities in~\cite{Meng}, and the new chained inequalities. The yellow
dashed lines, the green solid lines and the blue solid lines are for (\ref{eq18}), inequalities in~\cite{Meng}, and the new chained inequalities, respectively.}
\end{figure}

We also plot the relationship between $\mathcal{V}_{n}$ and $n$ for (\ref{eq18}), the chained steering inequalities in ~\cite{Meng}, and the new chained steering inequalities in Fig.~\ref{fig1}. We can see  that the new  chained inequality has the lowest visibility for  $n=4,6,10$.

\section{Visibility of quantum violation with the generalized Werner states}\label{BI}
Now we  investigate the visibility for the generalized Werner states
\begin{equation}\label{eq9-0}
\varrho=V|\psi(\theta)\rangle\langle\psi(\theta)|+(1-V)\frac{I_4}{4},
\end{equation}
where
\begin{equation}
|\psi(\theta)\rangle=\cos{\theta}|00\rangle+\sin{\theta}|11\rangle.
\end{equation}
Let Bob choose directions along the vertices of Platonic solids (as in Figure 2 of \cite{wisemanNature}) for (\ref{eq18}), and directions (\ref{eq15}), (\ref{eq17}) and (\ref{eq18-0}) for (\ref{eq1}), then one can find their corresponding threshold values of violation.

\begin{figure*}[tbp]
\subfigure[]{\includegraphics[width=53mm]{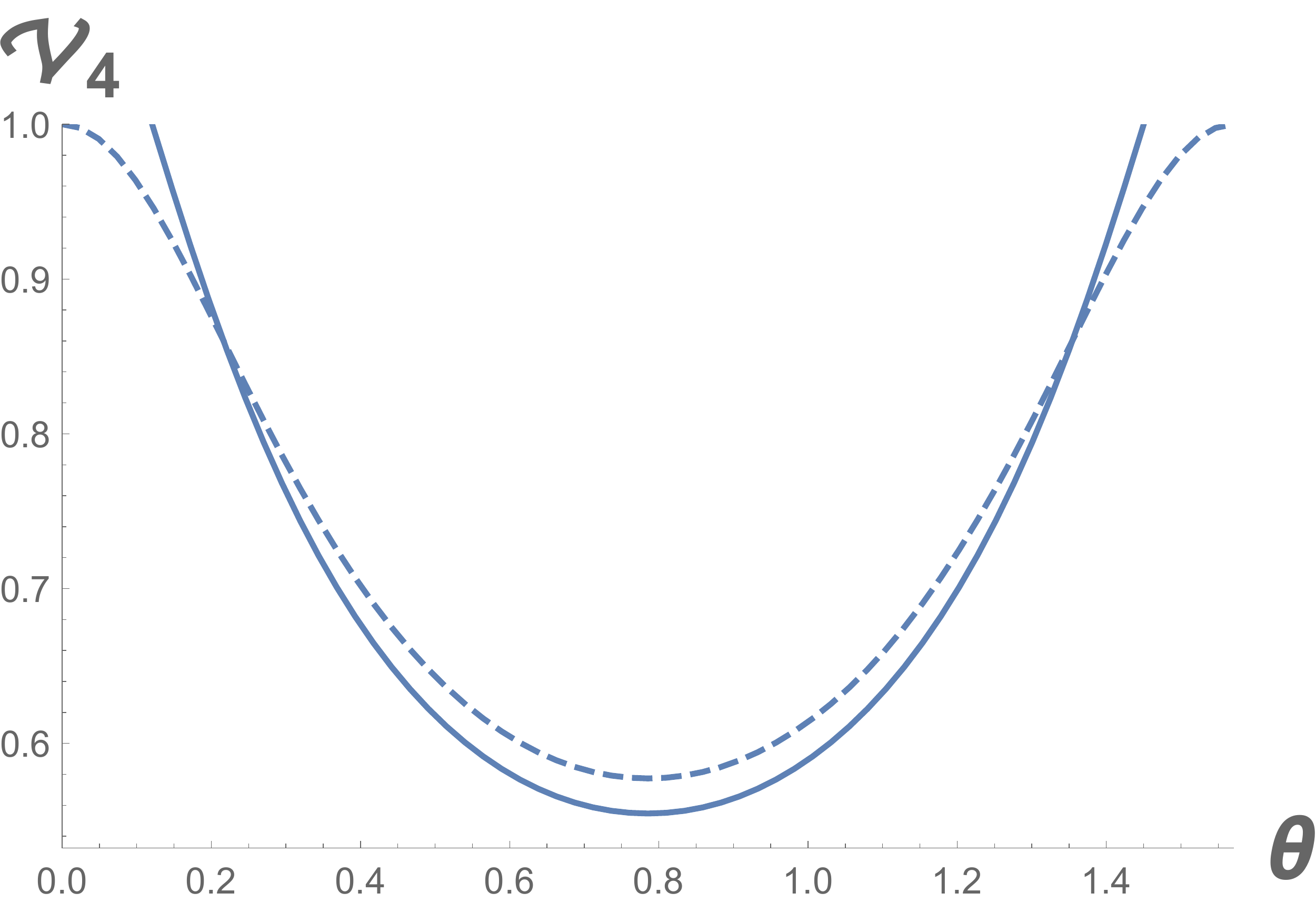}\label{fig2}}\hspace{9mm}
\subfigure[]{\includegraphics[width=53mm]{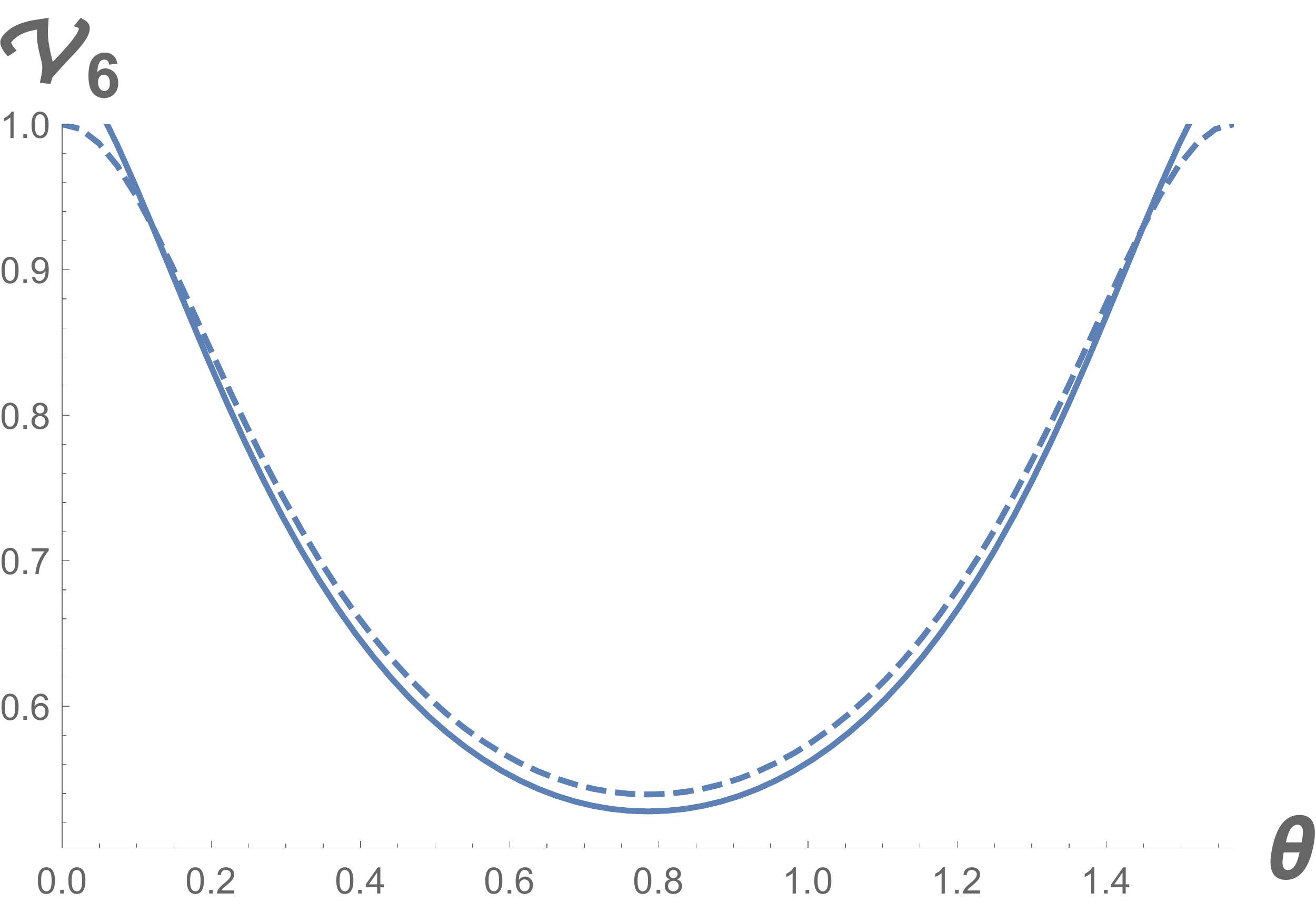}\label{fig3}}\hspace{9mm}
\subfigure[]{\includegraphics[width=53mm]{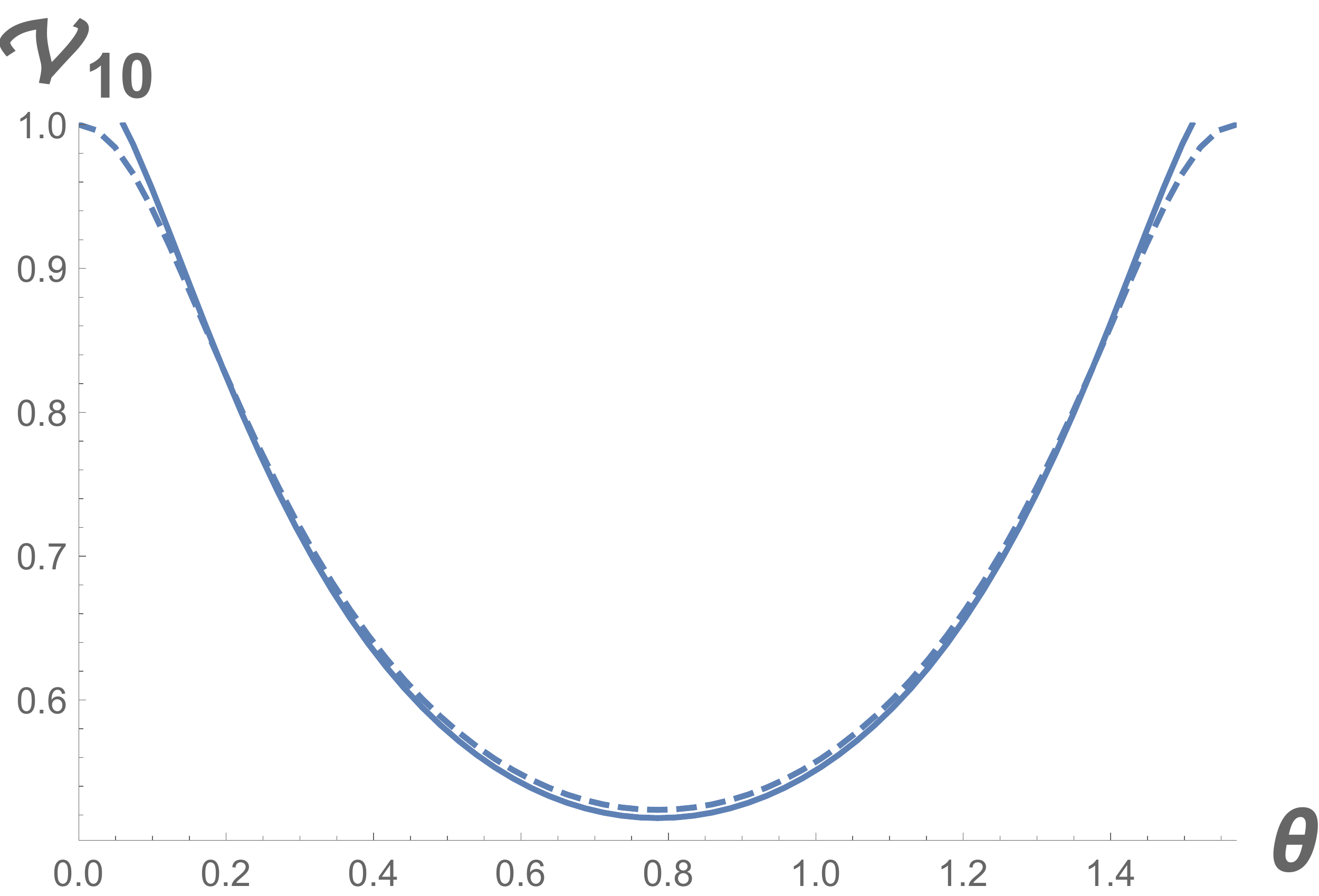}\label{fig4}}\hspace{9mm}
  \caption{For $n=4,6,10$, the blue solid curve and  dashed lines are the threshold value over the parameter $\theta\in[0,\pi/2]$ for the chained steering inequality and the linear steering inequality, respectively.}\label{figAll}
\end{figure*}

The results are summarized in Table~\ref{table1} and Fig.~\ref{figAll}. Before ending this section, we have a few observations.

\emph{Observation 1.---}The smallest threshold value is attained at $\theta=\pi/4$, reducing to the results in previous sections where the standard Werner state is considered.

\emph{Observation 2.---}For inequality (\ref{eq18}), one can obtain analytically
\begin{align}
  &Q_4=\frac{\sqrt{1+2\sin^2{2\theta}}}{\sqrt{3}},~~V_4=\frac{1}{\sqrt{1+2\sin^2{2\theta}}},\\
   &Q_6=\frac{1}{6}\left(1+\sqrt{5}\sqrt{1+4\sin^2{2\theta}}\right),\nonumber\\
   & V_6=\frac{1+\sqrt{5}}{1+\sqrt{5}\sqrt{1+4\sin^2{2\theta}}},\\
   &Q_{10}=\frac{1}{10}(1+\sqrt{5+4\sin^2{2\theta}}+2\sqrt{1+8\sin^2{2\theta}}),\nonumber\\
   & V_{10}=\frac{3+\sqrt{5}}{1+\sqrt{5+4\sin^2{2\theta}}+2\sqrt{1+8\sin^2{2\theta}}}.
\end{align}

\emph{Observation 3.---}For the new chained steering inequalities, i.e., inequality (\ref{eq1}) with directions (\ref{eq15}), (\ref{eq17}) and (\ref{eq18-0}),
the $\mathcal{V}_n\leq1$ if and only if $\theta\in[\theta_{n},\theta_{n}']$, $n=4,6,10,$ where
\begin{equation}
\begin{split}
&\theta_4\approx0.120673\approx\frac{59\pi}{1536},\\
 &\theta_4'=\frac{\pi}{2}-\theta_4\approx1.45012\approx\frac{709\pi}{1536},\\
 &\theta_6\approx0.0605856\approx\frac{3857\pi}{200000},\\
 &\theta_6'=\frac{\pi}{2}-\theta_6\approx1.51021\approx\frac{96143 \pi}{200000},\\
 &\theta_{10}\approx 0.0610922\approx\frac{15557 \pi}{800000},\\
 &\theta_{10}'=\frac{\pi}{2}-\theta_{10}\approx 1.5097\approx\frac{384443\pi}{800000}.
 \end{split}
 \end{equation}

\emph{Observation 4.---}The visibility of the new chained steering inequalities is lower than that of inequality (\ref{eq18}) when
\begin{equation}
\begin{split}
&\theta\in[0.25981,1.35482]~~{\rm for}~~n=4,\\
&\theta\in[0.121383,1.44941]~~{\rm for}~~n=6,\\
&\theta\in[0.193389,1.37741]~~{\rm for}~~n=10.
\end{split}
\end{equation}

\begin{table}[!h]
\caption{\label{table1} Region of $\theta$ in which $\mathcal{V}_n\leq 1$, $n=4,6,10.$}
\begin{ruledtabular}
\begin{tabular}{cccc}
$\theta\in[\theta_n,\theta'_n]$  & $n=4$ & $n=6$ & $n=10$\\
  \hline
     Inequality (\ref{eq18})  & $[0,\frac{\pi}{2}]$ & $[0,\frac{\pi}{2}]$ & $[0,\frac{\pi}{2}]$\\
  The new inequality & $[\frac{59\pi}{1536},\frac{709\pi}{1536}]$ & $[\frac{3857\pi}{200000},\frac{96143 \pi}{200000}]$ & $[\frac{15557 \pi}{800000},\frac{384443\pi}{800000}]$
\end{tabular}
\end{ruledtabular}
\end{table}

\section{Conclusions}\label{conclusion}

We have constructed new $n$-setting chained steering inequalities with the lowest visibility compared with the linear steering inequalities and the chained steering inequalities in~{\cite{Meng}}.  For the generalized Werner states, it has been found that when  parameter $\theta$ of the states lies in some neighborhood of $\frac{\pi}{4}$, the new chained steering inequalities always have lower visibility and thus more robust against noise than the $n$-setting  steering inequalities.
Subsequently, we shall try to construct optimal steering inequalities that have genuine minimal threshold values of quantum violation.

\begin{acknowledgments}
H.X.M. is supported by Project funded by China Post-doctoral Science Foundation (No. 2018M631726). C.L.R. is supported by National key research and development program (No. 2017YFA0305200), the Youth Innovation Promotion Association (CAS) (No. 2015317), the National Natural Science Foundation of China (No. 11605205), the Natural Science Foundation of Chong Qing (No. cstc2015jcyjA00021), the project sponsored by SRF for ROCS-SEM (No. Y51Z030W10), the fund of CAS Key Laboratory of Quantum Information. J.L.C. is supported by National Natural Science Foundations of China (Grant No. 11475089).
\end{acknowledgments}

\end{document}